\documentclass[aps,prd,
twocolumn,epsf,nofootinbib]{revtex4}
\usepackage{amsfonts,latexsym,eucal,amsmath,amssymb,times,mathrsfs}
\usepackage[dvips]{graphicx}

\newcommand{\bfk}{\mbox{\scriptsize\boldmath{$k$}}}
\newcommand{\bm}[1]{\hbox{\boldmath{$#1$}}}

\newcommand{\Mp}{M_{\rm pl}}
\newcommand{\dd}{{\rm d}}

\begin{document}

\thispagestyle{empty}


\title{IR divergence does not affect the gauge-invariant curvature perturbation}
\date{\today}
\author{Yuko Urakawa$^{1}$}
\author{Takahiro Tanaka$^{2}$}
\address{\,\\ \,\\
$^{1}$ Department of Physics, Waseda University,
Ohkubo 3-4-1, Shinjuku, Tokyo 169-8555, Japan\\
$^{2}$ Yukawa Institute for Theoretical Physics, Kyoto university,
  Kyoto, 606-8502, Japan}

\preprint{2010-**-**, WU-AP/***/**, hep-th/1007.0468}


\begin{abstract}
We address the infrared(IR) divergence problem during inflation that
appears in the loop corrections to the primordial
perturbations. In our previous paper, we claimed that, 
at least in single field models, the IR
divergence is originating from the gauge artifact. 
Namely, diverging IR corrections should not appear in 
genuine gauge-invariant observables. 
We propose here one simple but explicit example of such gauge-invariant
 quantities. 
Then, we explicitly calculate such a quantity 
to find that the IR divergence is absent 
at the leading order in the slow-roll approximation 
for the usual scale invariant vacuum state. 
At the same time we notice that there is a 
subtle issue on the gauge-invariance 
in how to specify the initial vacuum state. 
\end{abstract}


\maketitle


\section{Introduction}
The precise measurements of the primordial
fluctuation provide us with valuable information of the early universe. 
It is widely accepted that the primordial
fluctuation originates from the quantum fluctuation of the inflaton
field. 
In the last
decades, it has been recognized that the perturbation theory in
the inflationary universe might break
down because of the infrared(IR) divergence from loop
corrections~\cite{IRdivergence}. (For a recent review, see Ref.~\cite{Seery:2010kh}.) 
During inflation, massless fields are known to
yield the scale invariant spectrum $P(k) \propto 1/k^3$ at linear
order. These fields contribute to the one-loop diagram with the four point
interaction as $\int \dd^3\!k/k^3$, which leads to the logarithmic
divergence. 

It has been an issue of debate whether the IR divergences are 
physical or not. 
If it were really physical, there might be a
possibility that loop corrections are observable. 
The possibility of secular growth of IR contributions 
has also been studied motivated as a possible explanation 
of the smallness of the cosmological constant~\cite{CC}. 
However, before we discuss implications of the IR effects, 
we need to carefully examine whether the reported IR divergences
are not due to careless treatment. (See also Ref.~\cite{ReplyCC})

In our previous work~\cite{IRsingle}, we pointed out the presence
of gauge degrees of freedom in the frequently used gauges
such as the comoving gauge and the flat gauge, 
and these gauge degrees of freedom are responsible for the IR 
divergences. 
There, we have shown that, if we 
fix the residual gauge degrees of freedom, the IR divergences 
automatically disappear in the single field model. 
(Multi-field case was discussed in a separate paper~\cite{IRmulti}.) 

In this paper, we reexamine our previous argument. 
If the IR divergences are really 
due to the residual gauge degrees of freedom, 
those divergences should disappear if we evaluate genuine
gauge-invariant 
quantities even though we do not fix the residual gauge. 
Hence, the demonstration of the genuine gauge-invariance of 
computed results would be necessary to obtain reliable 
predictions, which are to be compared with observations. 
In this brief report we demonstrate more explicitly that the IR divergence in the
curvature perturbation can be removed by looking at such 
genuine gauge-invariant quantities. 
In doing so, we shall notice that the choice 
of initial vacuum state is restricted. 
In the slow roll limit, we find that a natural vacuum state 
is surprisingly limited to the Bunch-Davies vacuum 
state as far as we consider Gaussian vacuum state for 
the inflaton perturbation in the flat slicing at the initial time.  


\section{Basic equations}
We consider the single field inflation model
 whose action takes the form 
\begin{eqnarray}
 S = \frac{\Mp^2}{2} \int \sqrt{-g}~ [R - g^{\mu\nu}\phi_{,\mu} \phi_{,\nu} 
   - 2 V(\phi) ] \dd^4x~, 
\end{eqnarray}
where $\Mp$ is the Planck mass and the scalar field $\Phi$ was 
rescaled so as to be non-dimensional as in Ref.~\cite{IRsingle}. 
Substituting the metric form 
\begin{eqnarray}
 \dd s^2 = - N^2 \dd t^2  + h_{ij} (\dd x^i + N^i \dd t) (\dd x^j + N^j
  \dd t)~, 
\end{eqnarray}
the action becomes
\begin{eqnarray}
 S&\!=&\!\frac{\Mp^2}{2} \int\! \sqrt{h} \Bigl[ N ~^s\!R - 2 N
  V(\phi) + \frac{1}{N} (E_{ij} E^{ij} - E^2) 
 \cr && \qquad \quad
 + \frac{1}{N} ( \dot{\phi}
  - N^i \partial_i \phi )^2 - N h^{ij} \partial_i \phi \partial_j \phi \Bigr] \dd^4x~,
\end{eqnarray}
where $E_{ij}$ and $E$ are defined by 
\begin{eqnarray}
 E_{ij} = \frac{1}{2} \left( \dot{h}_{ij} - D_i N_j
 - D_j N_i \right), \quad E = h^{ij} E_{ij} ~.
\end{eqnarray}

In this paper we work in the comoving gauge, defined by $\delta
\phi=0$. 
We decompose the spatial metric as
\begin{eqnarray}
  h_{ij} = e^{2 (\rho + \zeta) }  \gamma_{ij}
 =  e^{2 (\rho + \zeta) } \left[ e^{\delta \gamma} \right]_{ij}~, 
\label{Cond:comoving} 
\end{eqnarray}
where $a:=e^{\rho}$ is the background scale factor, 
and ${\rm tr}[ \delta \gamma ]=1$. 
Using the degrees of freedom of choosing the 
spatial coordinates, 
we further impose the gauge conditions $\partial^i \delta
\gamma_{ij}=0$. 

Varying the action with respect to $N$ and $N^i$, 
we obtain the Hamiltonian and momentum constraints as
\begin{eqnarray}
 && ^s\! R - 2 V -  N^{-2}  (E^{ij} E_{ij} - E^2 )  
 - N^{-2} \dot{\phi}^2  = 0~, \label{Eq:Hconst} \\
 && D_j \left[ N^{-1} \left( {E^j}_i - {\delta^j}_i E \right) \right] =
  0~. \label{Eq:Mconst}
\end{eqnarray}
For convenience, we factorize the scale factor from the metric as
\begin{eqnarray*}
 \dd s^2=e^{2\rho}  [ - (N^2 - \check{N}_i \check{N}^i) \dd \eta^2
 + 2 \check{N}_i \dd \eta \dd x^i + \check{h}_{ij} \dd x^i \dd x^j
		     ],
\end{eqnarray*}  
where we have defined
$\check{h}_{ij}:= e^{- 2\rho}h_{ij}=e^{2\zeta} \gamma_{ij}$,
$N_i=e^{\rho} \check{N}_i$, and 
$\check{N}^i := \check{h}^{ij} \check{N}_i = e^\rho N^i$.
Expanding the perturbations, 
${\cal Q}=\delta N(:=N-1), \check{N}_i, \zeta$, and 
$\delta \gamma_{ij}$ as
${\cal Q} = {\cal Q}_1 +  \tfrac{1}{2} {\cal Q}_2 + \cdots$,
the zeroth-order Hamiltonian constraint equation yields the
background Friedmann equation:
\begin{eqnarray}
 6 \rho'\,^2 = \phi'\,^2 + 2 \check{V}(\phi)~, \label{Eq:Friedmann}
\end{eqnarray}
where a prime ``$~{}'~$'' denotes the differentiation 
with respect to the conformal time
$\eta$, and $\check{V}(\phi) := e^{2\rho} V(\phi)$.
The constraint equations at the linear order are obtained as
\begin{eqnarray}
 && \check{V} \delta N_1 - 3 \rho' \zeta'_1 
   + \partial^2  \zeta_1 + \rho' \partial^i \check{N}_{i,1} =0~,
 \label{Eq:Hconst1}\\
 && 4 \partial_i \left( \rho' \delta N_1 - \zeta'_1 \right)
    - \partial^2 \check{N}_{i,1} 
 + \partial_i \partial^j \check{N}_{j,1}=0~. \label{Eq:Mconst1}
\end{eqnarray}
The higher-order constraints
can be obtained similarly.

\section{Origin of IR divergence}
\subsection{Influence of boundary conditions}
Solving the constraints
(\ref{Eq:Hconst}) and (\ref{Eq:Mconst}), we can express
$\delta N$ and $N_i$ in terms of the curvature perturbation $\zeta$.
Here we stress that the constraints (\ref{Eq:Hconst1}) and
(\ref{Eq:Mconst1}) are elliptic-type equations, 
which require boundary conditions to solve. 
Below we study the relation between the residual
gauge degrees of freedom and the integration constants appearing 
in the general solution of Eqs.~(\ref{Eq:Hconst}) and
(\ref{Eq:Mconst}). 
For illustrative purpose,
we consider the linear order, in which constraints are given by
Eqs.~(\ref{Eq:Hconst1}) and (\ref{Eq:Mconst1}), but the extension to
higher orders proceeds in a similar manner. 

Eliminating $\delta N_1$ from 
Eqs.~(\ref{Eq:Hconst1}) and (\ref{Eq:Mconst1}), we obtain
\begin{align}
 &\left(1- \frac{4\rho'\,^2}{\check{V}} \right) \partial_i \partial^j
  \check{N}_{j,1} - \partial^2  \check{N}_{i,1} \cr
 &\qquad \qquad \qquad + 2 \frac{\phi'\,^2}{\check{V}} \partial_i \zeta'_1
 - 4 \frac{\rho'}{\check{V}} \partial_i \partial^2 \zeta_1=0~.
 \label{Eq:constHM1}
\end{align}
Taking the divergence of Eq.~(\ref{Eq:constHM1}), the longitudinal part
of $\check{N}_{i,1}$ is solved as
\begin{eqnarray}
 \partial^i \check{N}_{i,1}(x) = \frac{\phi'\,^2}{2\rho'\,^2} \zeta_1'(x)
 - \frac{1}{\rho'} \partial^2 \zeta_1(x) 
 + F_1(x)~, \label{Sol:cNi1/L}
\end{eqnarray}
where $F_1(x)$ is an arbitrary solution of the Laplace equation. 
Using Eqs.~(\ref{Eq:constHM1}) and
(\ref{Sol:cNi1/L}),
$N_{i,1}$ is integrated to give
\begin{align}
 \check{N}_{i,1}(x) &= \partial_i 
 \left( \frac{\phi'\,^2}{2\rho'\,^2} \partial^{-2} \zeta_1'(x)
 - \frac{1}{\rho'}  \zeta_1(x) \right)
 \cr & \quad 
 + \left(1- \frac{4\rho'\,^2}{\check{V}} \right) \partial_i
 \partial^{-2} F_1(x)
 +G_{i}(x)~, \label{Sol:cNi1}
\end{align}
where $G_{i}(x)$ is an arbitrary vector that 
satisfies $ \partial^2 G_{i}(x) =0$. Here we assumed 
that the inverse Laplacian $\partial^{-2}$ is defined uniquely.  
Comparing the
divergence of Eq.~(\ref{Sol:cNi1}) with Eq.~(\ref{Sol:cNi1/L}),
we find 
$F_1(x) = (\check{V}/4 \rho'\,^2) \partial^i G_{i}(x)$.
Substituting Eq.~(\ref{Sol:cNi1/L}) into Eq.~(\ref{Eq:Mconst1}), the
lapse function is easily obtained.
To conclude, the lapse function and the shift
vector are given by 
\begin{align}
 & \delta N_1 = \frac{1}{\rho'} \left( \zeta_1' - \frac{1}{4}
  \partial^i G_{i}  \right)~,  \label{Exp:N1} \\
 & \check{N}_{i,1}= \partial_i 
 \left( \frac{\phi'\,^2}{2\rho'\,^2} \partial^{-2} \zeta_1'
 - \frac{1}{\rho'}  \zeta_1 \right) \cr
& \qquad \qquad
 - \frac{1}{4} \left( 1 + \frac{\phi'\,^2}{2 \rho'\,^2} \right) \partial_i
 \partial^{-2} \partial^j G_{j}
 +G_{i}~. \label{Exp:cNi1}
\end{align}
Introduction of a vector $G_i(x)$ also modifies the evolution equations of
the curvature perturbation $\zeta$ and the transverse-traceless
perturbation 
$\delta \gamma_{ij}$. 
Substituting Eqs.~(\ref{Exp:N1}) and (\ref{Exp:cNi1}) into Einstein 
equations, we obtain
\begin{eqnarray}
 \bigl( \zeta_1 - f_1 \bigr)'' + 2 (\ln z)' \bigl( \zeta_1 - f_1
  \bigr)' - \partial^2 \bigl( \zeta_1 - f_1 \bigr)=0~, \label{Eq:tildezeta0}
\end{eqnarray}
where we have defined $z:= e^\rho \phi'/\rho'$ and
$f_1(x):= \frac{1}{4} \int \dd \eta\, \partial^i G_{i}(x)$, which implies
$\partial^2 f_1(x)=0$.
While the evolution equation of $\delta \gamma_{ij}$
is obtained as 
\begin{align}
 &  \!\!\delta \gamma_{ij,1}'' + 2 \rho' \delta \gamma_{ij,1}' - \partial^2 \delta
  \gamma_{ij,1} \cr
 & \quad -  \left(\phi' \over \rho' \right)^2 \partial_i 
 \partial_j (\partial^{-2} \partial^2 - 1) \zeta_1   + \frac{1}{2\rho'}
 \partial_i \partial_j \partial^k  G_{k} \cr
 & \quad
 - e^{-2 \rho} \partial_{\eta} \bigl\{ e^{2\rho} ( \partial_i G_{j} + \partial_j
 G_{i}  \cr&\qquad \qquad
 - \tfrac{1}{2} \partial_i \partial_j \partial^{-2} \partial^k
 G_{k} - \tfrac{1}{2} \partial^k G_{k} \delta_{ij} ) \bigr\}
 =0~.
 \label{Eq:Einsteinij/tl3}
\end{align}
This equation is consistent with the transverse traceless condition 
of $\delta\gamma_{ij}$. Namely, the
trace and divergence of Eq.~(\ref{Eq:Einsteinij/tl3}) 
are automatically satisfied. These evolution equations
of $\zeta_1$ and $\delta \gamma_{ij,1}$ 
reduce to the ordinary well-known linear perturbation equations
by simply setting $G_i=0$.

When we consider the universe with infinite volume, 
the solution of $\delta N_1$ and $\check N_{i,1}$ would be specified uniquely by 
the requirement that all quantities are regular at the spatial infinity. 
In this case, we have not room to introduce $G_{i}(x)$. 
In contrast, when we concentrate on 
a finite region of the universe, 
a variety of solutions are allowed especially due to the presence 
of homogeneous solutions of Laplace equation. 
In this case, as presented by
the first term in the second line of Eq.~(\ref{Eq:Einsteinij/tl3}), 
it also happens that the 
evolution equation for the transverse traceless mode 
$\delta\gamma_{ij}$ is contaminated by the longitudinal mode $\zeta$, 
even at the linear order. 

\subsection{Gauge degrees of freedom}
The ambiguity in the choice of the vector $G_{i}$ 
indicates the presence of residual gauge degrees of freedom.
Here, we show explicitly that $G_{i}(x)$ 
represents the residual gauge degrees of freedom that 
remain even after we specify the gauge by 
$\delta \phi=0$ and the conditions~(\ref{Cond:comoving}). 
Since the condition $\delta \phi=0$ completely fixes 
the temporal gauge, we are only allowed to change the spatial coordinates.
Under the change of the spatial coordinates $x^i\to x^i+\delta x^i$,
the metric functions transform as
\begin{align}
 & \tilde{\check{N}}_{i,1}(x) = \check{N}_{i,1}(x) - \delta x_i'~, 
\label{Trans:cNi1} 
\\
 & \tilde{\zeta}_1(x) = \zeta_1(x) - \tfrac{1}{3} \partial^i \delta
 x_i~, \label{Trans:zeta1} \\
 & \delta \tilde{\gamma}_{ij,1} = \delta \gamma_{ij,1} - \left( \partial_i \delta x_j + \partial_j \delta x_i 
 - \tfrac{2}{3} \partial^k \delta x_k \delta_{ij} \right)~, \label{Trans:gamma1}
\end{align}
where $\delta x_i := \delta_{ij} \delta x^j$.
We associate a tilde with the perturbed variables when we set 
$G_{i}\ne 0$ to discriminate them 
from those in the gauge with $G_{i} = 0$. 
Recalling that the lapse function remains unchanged, 
the comparison between Eqs.~(\ref{Exp:N1}) and (\ref{Trans:zeta1}) 
gives
\begin{eqnarray}
 \partial^i \delta x_i' = - (3/4) \partial^i  G_{i}(x)~. 
 \label{Exp:deltaxi1}
\end{eqnarray} 
Imposing the transverse condition of $\delta \gamma_{ij}$ in 
Eq.~(\ref{Trans:gamma1}) yield
\begin{eqnarray}
 \partial^2 \delta x_i = - (1/3) \partial_i \partial^j \delta x_j~. 
 \label{Exp:deltaxi2}
\end{eqnarray}
Comparing Eq.~(\ref{Exp:cNi1}) with Eq.~(\ref{Trans:cNi1}) 
together with Eqs.~(\ref{Exp:deltaxi1}) and (\ref{Exp:deltaxi2}), 
$\delta x_i$ is solved as
(See Ref.~\cite{IRgauge}.)
\begin{align}
 \delta x_i(x) &=   -   \int \hspace{-0.1cm} \dd \eta\, G_{i}(x) +
 \frac{1}{4} \int \hspace{-0.1cm} \dd \eta \partial_i \partial^{-2}
  \partial^j G_{j}(x) \cr &\qquad 
 +  \frac{1}{4} \int\hspace{-0.1cm} \frac{\dd \eta}{\rho'} \int \hspace{-0.1cm} \dd \eta \, \partial_i
 \partial^j G_{j}(x)  + H_{i}(\bm{x}) 
 \cr &\qquad  + 
 \int \frac{\dd \eta}{\rho'} \, \partial^2 H_{i}(\bm{x})~, \label{Exp:dxi}
\end{align}
where we introduced a vector $H_{i}(\bm{x})$ that satisfies
\begin{eqnarray}
 3 \partial^2 H_{i}(\bm{x}) + \partial_i \partial^j H_{j}(\bm{x})=0~.
\end{eqnarray}
Substituting Eq.~(\ref{Exp:dxi}) into Eqs.~(\ref{Trans:zeta1}) and 
(\ref{Trans:gamma1}), we find that 
spatial components of metric perturbation
transform as
\begin{eqnarray}
 \tilde{\zeta}_1 &\!=&\! \zeta_1 + \frac{1}{4}\hspace{-0.1cm}\int \hspace{-0.1cm} \dd \eta\,
 \partial^i G_{i} - \tfrac{1}{3} \partial^i H_{i}~. \\
 \delta \tilde{\gamma}_{ij,1} &\!=&\! \delta \gamma_{ij,1} + 
 \hspace{-0.1cm}\int \hspace{-0.1cm} \dd \eta \left\{ 2\partial_{(i} G_{j)}
  - \tfrac{1}{2} (\partial_i \partial_j \partial^{-2}+\delta_{ij})
 \partial^k G_{k}   \right\} \cr
&&  -\! \frac{1}{2}\hspace{-0.1cm} \int\hspace{-0.1cm}
 \frac{\dd \eta}{\rho'}\hspace{-0.1cm} \int\hspace{-0.1cm} \dd \eta \partial_i
 \partial_j \partial^k G_{k}\! - 2 \left( \partial_{(i} H_{j)}
 - \tfrac{1}{3} \partial^k H_{k} \delta_{ij} \right) \cr
&& - 2 \int \frac{\dd \eta}{\rho'} \partial^2 \partial_{(i}
 H_{j)}.
\end{eqnarray}

To remedy the IR divergence due to the gauge modes, in our previous
work~\cite{IRsingle, IRmulti},
we proposed to use the local gauge condition such that the effects from the
causally disconnected region are shut off. 
The local gauge condition can be achieved by 
choosing the function $G_i(x)$ appropriately. 
To remove the IR divergence explicitly, 
the residual gauge must be fixed so that 
$\tilde\zeta$ measures the deviation from the average value 
taken over a certain local region ${\cal O}$. 
In Ref.~\cite{IRsingle}, we have
shown that the IR corrections become finite 
by adopting this local gauge. 
In this gauge IR contributions are effectively 
cut off at the length scale of the size of ${\cal O}$.
However, one can change the choice of $\cal O$, 
which leads to a finite shift in the gauge function $G_{i}(x)$. 
As a result, the resulting correlation functions for 
$\tilde\zeta$ depend on the artificial choice of $\cal O$. 
This prescription will be sufficient just to show that 
the loop corrections are finite in some gauge, but 
it will not be sufficient to evaluate the size of 
loop corrections.  
In the present work, we give an example of 
genuine gauge-invariant quantities
that do not depend on how we fix the residual gauge degrees of freedom. 
The expected correlations in the CMB 
temperature fluctuations should be also such gauge-invariant quantities.

\section{Gauge invariance and IR regularity}
\subsection{Construction of gauge-invariant variables}
In this subsection, we construct an example of genuine gauge-invariant quantities. 
Since the temporal slicing is already fixed, 
it is sufficient to address the gauge-invariance 
under the residual gauge transformation of the spatial 
coordinates~(\ref{Exp:dxi}).
Genuine gauge-invariance is 
equivalent to complete gauge fixing. 
In this sense, if we give appropriate boundary conditions 
for the lapse and shift, the residual gauge degrees of 
freedom can be completely fixed. 
However, here arises a difficulty in fixing the gauge, because we need
to remove all arbitrariness regarding the choice of coordinates, such as the 
choice of the local region $\cal O$.  

When we consider the universe with infinite volume, the 
conventional gauge-invariant perturbation theory is formulated 
using particular combinations of
perturbed variables invariant under an arbitrary gauge
transformation. 
The construction of such gauge-invariant variables is possible 
thanks to the absence of the ambiguity in the inverse
Laplacian under the assumption that the perturbations remain finite at the spatial infinity. 
However, the same procedure does not work 
when we try to use only the information contained 
in the limited area ${\cal O}$.
Under this restriction, genuine gauge-invariant
quantities cannot be constructed by the combination of local quantities. 

Keeping these in
mind, we consider $n$-point functions 
of the scalar curvature
$^s\!R$. The scalar curvature $^s\!R$ does not remain invariant but 
transforms as a scalar quantity under the change of spatial
coordinates. If we could specify its $n$ arguments 
in a coordinate-independent manner, the $n$-point functions of $^s\! R$ 
would be gauge-invariant. 
This can be partly achieved by specifying the $n$ spatial points by 
the geodesic distances and the directional cosines from 
one reference point. Although we cannot specify the reference 
point in a coordinate independent manner, 
this gauge dependence would not matter as long as we are interested in 
the correlation functions for a quantum state that respects the spatial 
homogeneity and isotropy of the universe. 

As an example, we consider the two point function of $^s\!R$, 
whose arguments $X_{A}^i$ ($A=1,2$) are specified 
by solving the three-dimensional geodesic equation 
from $\lambda=0$ to $1$ 
with the initial ``velocity'' $\dd X_{A}^i(\lambda)/\dd \lambda|_{\lambda=0}=x_{A}^i$ 
starting with the origin.  
We denote $X_{A}^i(\lambda=1)$ simply by $X_{A}^i$, which can 
be perturbatively expanded as $X_{A}^i:=x_{A}^i+\delta x_{A}^i$. 
$x^i_{A}$ can be understood as the background value of the points. 
Then, the expectation value of a product of 
\begin{align}
\!\! ^g\! R(\eta,\,x_{A}^i)&:= {^s\! R} (\eta,\,x_A^i+\delta x_A^i) \cr
 &= \sum_{n=0}^\infty \frac{\delta x_A^{i_1} \cdots \delta x_A^{i_n}}{n!}
 \partial_{i_1} \cdots \partial_{i_n}\!{^s\!R}(\eta,\, x_A^i), \label{Def:Rg}
\end{align}
should be gauge-invariant. 

\subsection{IR regularity of genuine gauge-invariants}
The genuine gauge-invariant quantities we introduced 
should be finite in the local gauge, since the field itself 
is constructed to be free from IR divergence in this gauge by construction. 
However, since these quantities are really gauge-invariant, 
they should be finite
even if we send the size of ${\cal O}$ to
infinity. This limit is supposed to agree with the case when we
calculate them in the infinite volume~\cite{IRgauge}.
In the rest of this paper we study 
the regularity of the gauge-invariant two point function 
introduced above in the conventional global gauge~\cite{Maldacena:2002vr}, focusing on 
the loop of longitudinal modes at the 
lowest order in the slow-roll approximation.
The spatial curvature $^s\!R$ is given by
\begin{align}
 &^s\!R= -2 e^{- 2 (\rho + \zeta)}(2 \partial^2 \zeta+ \partial_i
 \zeta\partial^i \zeta)~. 
\end{align} 
We expand the gauge-invariant spatial curvature as 
$^g\! R={^g\!R}_1+ {^g\!R}_2+ \cdots$, 
in terms of the interaction picture field operator.
Here the subscript denotes the number of the operators. 
Then, one-loop contribution to the two point function 
starts with the quartic order, 
\begin{eqnarray}
 \langle {^g\!R} {^g\!R} \rangle_{4} :=
  \langle {^g\!R}_1 {^g\!R}_3 \rangle +  \langle {^g\!R}_2 {^g\!R}_2 \rangle
  + \langle {^g\!R}_3 {^g\!R}_1 \rangle~.
\end{eqnarray}
Using the fact that the IR divergence arises only
from the contraction between interaction picture fields without any
derivative, we keep only the terms that are possibly divergent. For
instance, the terms that include more than two interaction picture
fields with spatial or temporal derivatives do not yield
divergences.

In order to obtain the expression for $^s\!R$, we use the 
fact that in the flat slicing the interaction Hamiltonian is 
totally suppressed by the slow roll parameter~\cite{Maldacena:2002vr}. 
We therefore solve the non-linear evolution of perturbation 
in the flat gauge, and transform the results into 
the $\delta\phi=0$ gauge. Then, the contributions 
at the lowest order in slow-roll expansion to $\zeta$ 
arises only from the non-trivial gauge transformation given by~\cite{Maldacena:2002vr} 
\begin{equation}
 \zeta=\zeta_n+{1\over 2\rho'}\zeta'_n\zeta_n+{1\over
  2\rho'{}^2}\zeta''_n\zeta_n^2+\cdots\, , 
\end{equation}
where $\zeta_n:=-\rho'\varphi_I/\phi'$ and the 
abbreviated terms are higher order or irrelevant for the divergences. 
Here $\varphi_I$ is the interaction 
picture field corresponding to the inflaton perturbation 
in the flat gauge. 

Solving the spatial geodesic equation 
on a $\eta=$constant hypersurface, we obtain
$X^i=e^{-\zeta} x^i+\cdots$, where 
the terms with spatial differentiations were abbreviated  
since they do not contribute to the 
possible IR divergence in $ \langle {^g\!R} {^g\!R} \rangle_{4}$. 
After some manipulations, the possibly divergent terms
in $ \langle {^g\!R} {^g\!R} \rangle_{4}$  can be summed up as
\begin{eqnarray}
\langle {^g\!R}(x) {^g\!R}(y) \rangle_{4}
  &\!\!\propto\!\!& 
 \langle \zeta_n^2 \rangle 
\!\!\int\! {\dd} (\log k)
  \bigl[D^2u_{\bfk}(x) u^*_{\bfk}(y)
\cr &&
         +2Du_{\bfk}(x) Du^*_{\bfk}(y)
         +u_{\bfk}(x)D^2u^*_{\bfk}(y)
          \bigr]
\nonumber\\ &&
+(\mbox{c.c.})\,, \label{Exp:regularity}
\end{eqnarray}
where $u_{\bfk}(x)$ is the positive frequency function of
the field $\zeta_n$ multiplied by $k^{7/2}$, and 
$D$ is an operator 
defined by $D:=\eta\partial_\eta+x^i\partial_i+2$.  
Since $\langle\zeta_n^2\rangle$ is IR divergent, 
the regularity is maintained only when the integral in (\ref{Exp:regularity})
vanishes exactly. 
If we use the scale-invariant 
property of mode functions of the Bunch-Davies
vacuum, the derivative operator $D$ acting on 
$u_{\bfk}$ is replaced with $\partial_{\log k}$.
Then, this integral becomes total derivative and vanishes.
The usage of the invariant distance is requested also in
Ref.~\cite{Byrnes:2010yc}, where the $\delta N$ formalism is utilized. 
It is intriguing that, while the method is different, they have arrived at
the same conclusion as we have.

One may think it strange to require the 
scale-invariance, because genuine 
gauge-invariant correlation functions should 
be finite independent of the choice of initial vacuum state. 
This finiteness is guaranteed as long as we compute it 
in the local gauge. 
This gauge dependence stems from that of the 
initial vacuum state. 
In Ref.~\cite{IRsingle} we implicitly specified the initial vacuum 
state so that the interaction picture field is 
identical to the Heisenberg field on the initial surface. 
Namely, the interaction is assumed to be turned off 
before the initial time. 
Since the effect of the coordinate transformation 
on the perturbation variables is non-linear, 
this identification between the Heisenberg field 
and the interaction picture field has different 
meaning depending on the choice of gauge. 
Therefore, in order to obtain genuine gauge-invariant results, 
we need the principle to specify the vacuum state in a gauge-invariant 
manner. 
If we specify the initial state so that the interaction 
picture field is identical to the Heisenberg field 
at the initial time in the local gauge, the selected quantum 
state depends on the choice of the local volume ${\cal O}$.  
This situation is unsatisfactory. Moreover, such a choice of vacuum 
will violate the three-dimensional translational invariance. 
Hence, it is much favoured to specify the initial 
vacuum state in the global gauge. 
As we have seen above, at the lowest order of slow roll approximation, 
the scale-invariant Bunch-Davies vacuum state is such a vacuum state 
that is free from IR divergences. 
Since the origin of the possible IR divergence is 
confined to the initial vacuum state,  
we may say that, if genuine gauge-invariant 
correlation functions of fluctuations are free from divergence 
at the initial time, it remains so in the course of time evolution.  
However, the extension of a natural vacuum free from IR divergence  
to the higher order in slow-roll expansion 
does not seem trivial at all. 
In this brief report we have also neglected the transverse-traceless metric 
perturbation, which can participate in the IR divergence. 
These issues will be discussed in our future
publication~\cite{IRgauge}.

\acknowledgments
The discussions during the workshops YITP-T-09-05 and YITP-T-10-01 
at Yukawa Institute were
very useful to complete this work. YU and TT would like
to thank Arthur Hebecker for his valuable comments.
T.~T. is supported by
the JSPS through Grants No.\ 21244033. We also acknowledge the support
of the Grant-in-Aid for the Global COE Program ``The Next Generation of
Physics, Spun from Universality and Emergence'' and the Grant-in-Aid for
Scientific Research on Innovative Area No.\ 21111006 from the MEXT.

\end{document}